\documentclass{webofc}

\usepackage[varg]{txfonts}   
\usepackage{hyperref}
\usepackage{url}
\usepackage{amsmath}
\usepackage{subfigure}
\usepackage{physics}

\hypersetup{colorlinks=true,citecolor=blue,urlcolor=blue,linkcolor=blue}
\begin{document}
\title{Top-quark Yukawa coupling with a dimension-6 operator}
%
%

\author{\firstname{Ya-Juan} \lastname{Zheng}\inst{1}\fnsep\thanks{\email{yjzheng@iwate-u.ac.jp}} 
}

\institute{Faculty of Education, Iwate University, Morioka, Iwate 020-8550, Japan 
          }

\abstract{We extend the standard top-quark Yukawa coupling with a dimension-6 operator in order to accommodate a CP violating complex phase with manifest gauge invariance. This leads to a new $ttHH$ contact interaction, along with many Goldstone boson couplings. We investigate the impact of the new interactions on a muon collider process $\mu^-\mu^+\to \nu_\mu\bar{\nu}_\mu t\bar{t}H$ compared with the standard dimension-4 top-Yukawa coupling. The unitarity bounds on the coefficient of the new physics operator is obtained from $W^-_LW^+_L$ and $HH$ initiated processes. 
}
\maketitle
\begin{figure}[t]
\centering
{
\includegraphics[width=5cm,clip]{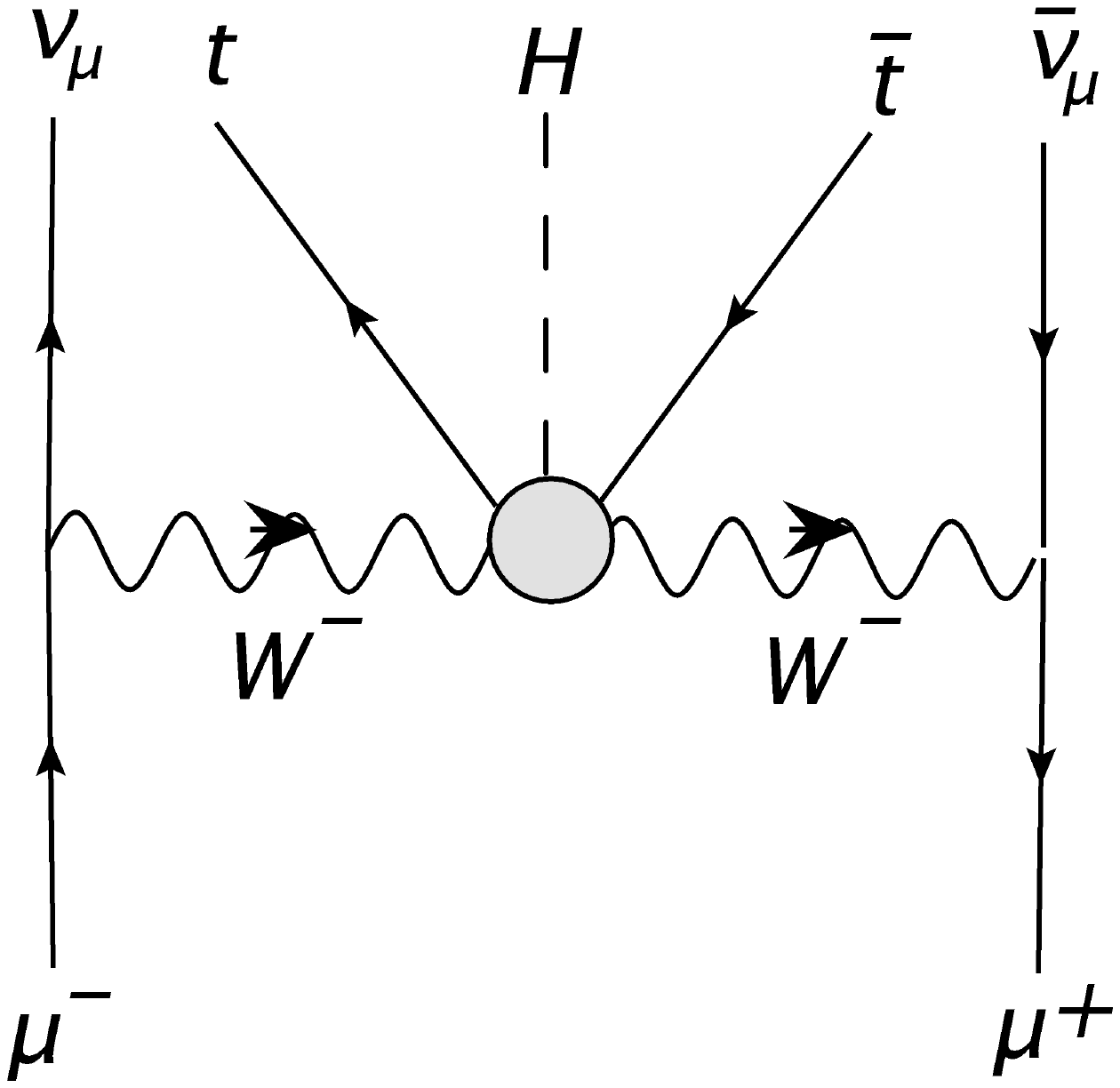}}
\caption{Weak boson fusion subdiagrams contributing to the process $\nu_\mu\bar{\nu}_\mu t\bar{t}H$.}
\label{fig:Feynman}       
\end{figure}

%
\label{intro}
The observed baryon asymmetry of the universe (BAU) requires a new source of CP violation among elementary particles. We consider the top-quark Yukawa coupling with a CP-violating phase~\cite{Sakharov:1988vdp,Kuzmin:1985mm}. 
This can be realized by introducing a complex Yukawa form:
\begin{eqnarray}
    {\cal L}^{ttH}_{\rm complex}=
    -gH\bar{t}(
    \cos\xi+i\gamma_5\sin\xi
    )t,
    \label{eq:complex}
\end{eqnarray}
where the phase $\xi~(-\pi\leq\xi\leq\pi)$ gives CP violation when $\sin\xi\neq0$
\footnote{In the case where $\sin\xi=1$, a CP-odd $Htt$ interaction can still induce CP violation through amplitudes with a CP-even $HVV$ interactions such as $VV\to t\bar{t}H,(V=W^\pm,Z)$. In contrast, in processes such as $gg\to t\bar{t}H$ and $\gamma\gamma\to t\bar{t}H$, no CP violating effects arise when $\sin\xi=1$, in the tree-level, as the amplitudes with only purely CP-odd $H$ couplings preserve the CP symmetry.}. 
We find that at high energies, we need a gauge invariant form of the interaction, by introducing a dimension-6 operator~\cite{Zhang:1994fb,Barger:2023wbg,Cassidy:2023lwd,Bar-Shalom:2024dav},
\begin{eqnarray}
    {\cal L}_{\rm SMEFT}^{ttH}={\cal L}_{\rm SM}^{ttH}+
    \left\{\frac{\lambda}{\Lambda^2}\left(Q^\dagger\phi t_R\right)\left(\phi^\dagger\phi-\frac{v^2}{2}\right)+{\rm h.c.}\right\},
    \label{eq:LSMEFT}
\end{eqnarray}
with
$\lambda$ a complex number which measures deviation from the SM,  $\Lambda$ is the new physics scale, and
\begin{eqnarray}
Q=
\begin{pmatrix}
t_L \\
b_L
\end{pmatrix},
\quad\quad
\phi=
\begin{pmatrix}
\frac{v+H+i\pi^0}{\sqrt{2}} \\
i\pi^-
\end{pmatrix}.
\end{eqnarray}
A more generic higher dimensional operator case has been discussed e.g.\,in~\cite{Brod:2022bww,Bar-Shalom:2024dav}.
We study consequences of the CP-violating top-Higgs coupling in the muon collider process $\mu^-\mu^+\to \bar{\nu}_{\mu}\nu_\mu\bar{t}tH$ with the vector boson fusion production diagrams shown in Fig.~\ref{fig:Feynman}.
After expansion, the Yukawa Lagrangian becomes 
\begin{eqnarray}
    {\cal L}_{\rm SMEFT}^{ttH}
    &=&
    -\sqrt{2}g_{\rm SM}\left(Q^\dagger\phi t_R\right)+\frac{\lambda}{\Lambda^2}\left(Q^\dagger\phi t_R\right)\left(\phi^\dagger\phi-\frac{v^2}{2}\right)+{\rm h.c.},
    \nonumber\\
    &=&-m_tt_L^\dagger t_R
    -g_{\rm SM}\left[\left(H+i\pi^0\right)t_L^\dagger+i\sqrt{2}\pi^-b_L^\dagger\right]t_R
    \nonumber\\
    &&+(g_{\rm SM}-ge^{i\xi})\left\{Ht_L^\dagger t_R+\frac{H}{v}
\left[\left(H+i\pi^0\right)t_L^\dagger+i\sqrt{2}\pi^-b_L^\dagger\right]t_R
    \right\}
    \nonumber\\
    &&+(g_{\rm SM}-ge^{i\xi})
    \Bigg\{
    \left(
    \frac{H^2+(\pi^0)^2}{2v}+\frac{\pi^+\pi^-}{v}
    \right)t_L^\dagger t_R
    \nonumber\\
    &&+\frac{H^2+(\pi^0)^2+2\pi^+\pi^-}{2v^2}
    \left[
    \left(
    H+i\pi^0
    \right)
t_L^\dagger+i\sqrt{2}\pi^-b_L^\dagger
    \right]t_R
    \Bigg\}+{\rm h.c.},
\label{eq:LSMEFT}
\end{eqnarray}
with $g_{\rm SM}=\frac{m_t}{v}$ and $g_{\rm SM}-ge^{i\xi}=\frac{\lambda v^2}{\sqrt{2}\Lambda^2}$. 
There are new couplings, $ttHH$ and $ttHHH$, proportional to $g_{\rm SM}-ge^{i\xi}$, which will lead to different CPV parameter $\xi$ dependence compared with the complex Yukawa parametrization Eq.\eqref{eq:complex}.
In the phenomenogical study, we will show how different the prediction can be at certain kinematical region, especially at high energies. 

\begin{figure}[b]
\centering
\subfigure[]{
\includegraphics[width=5cm,clip]{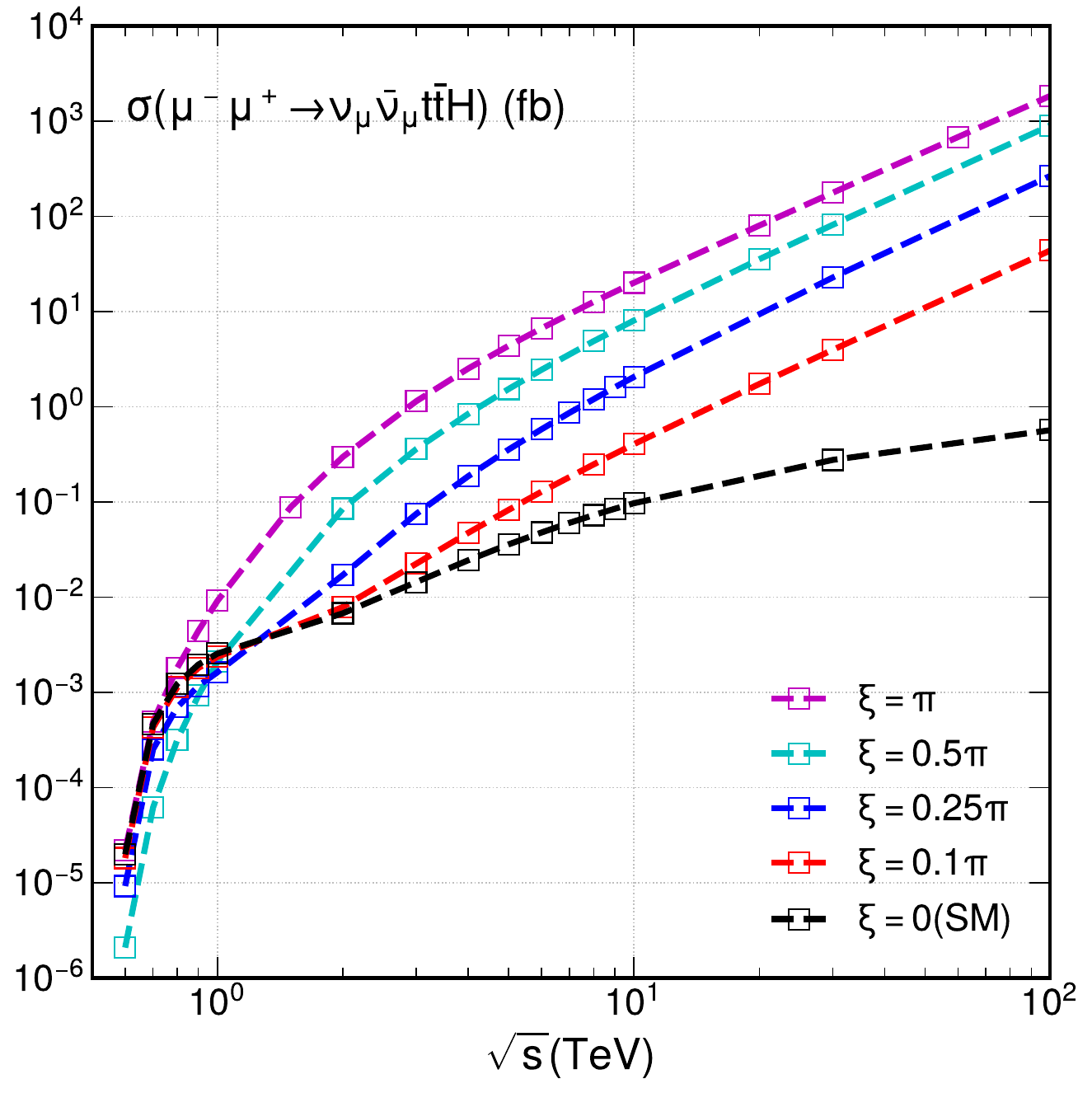}}
\subfigure[]{
\includegraphics[width=5cm,clip]{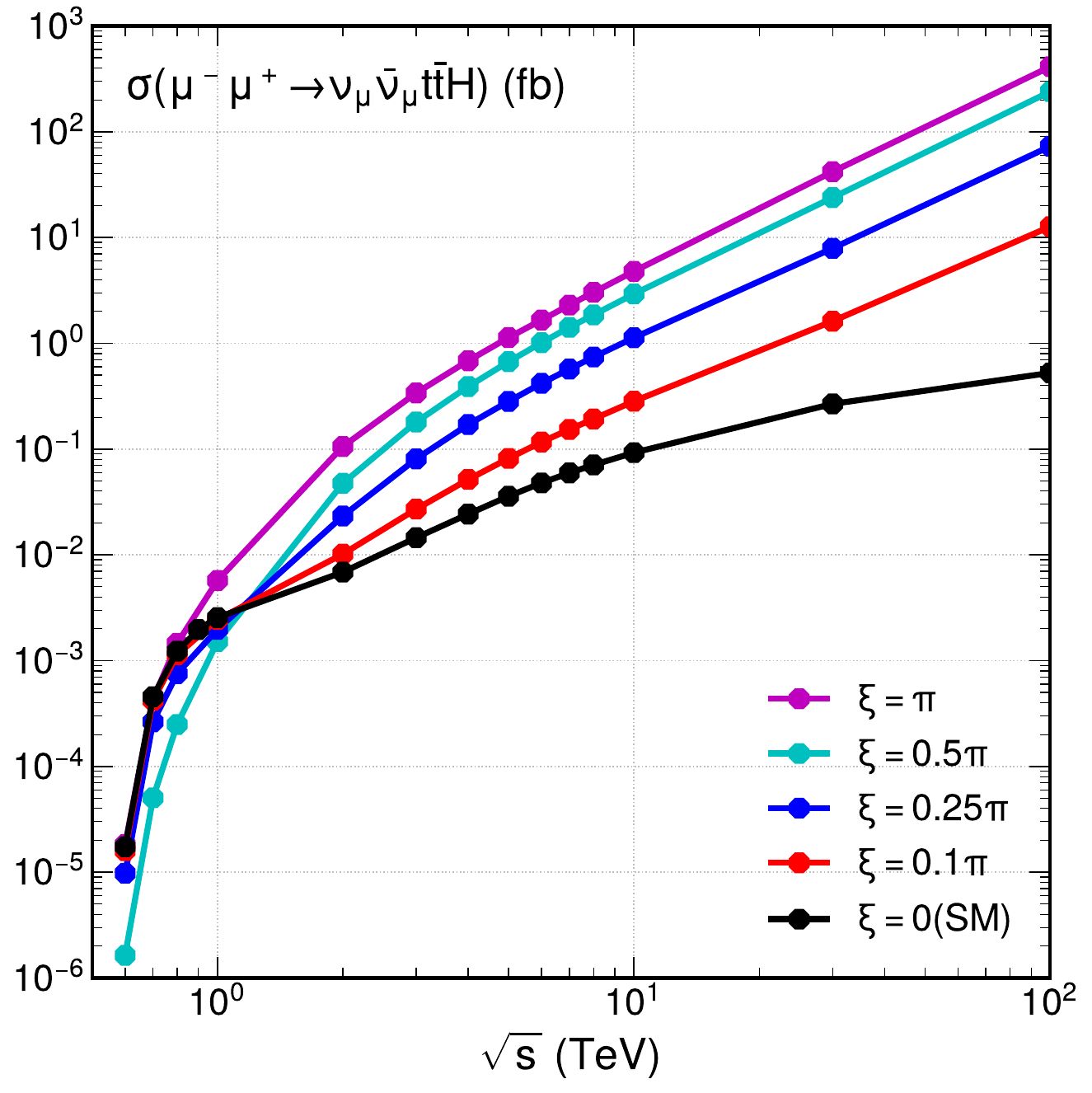}}
\caption{Total cross section of $\nu_\mu\bar{\nu}_\mu t\bar{t}H$ production at a muon collider with $\sqrt{s}$ dependence at $\xi=0, 0.1\pi,0.25\pi,0.5\pi,\pi$ (a) in the complex Yukawa model, (b) in the SMEFT model.}
\label{fig-1}       
\end{figure}

Comparing the complex Yukawa model and the one with an extra dimension-6 operator, the $ttHH$ coupling Lagrangian
\begin{eqnarray}
    {\cal L}_{ttHH}=\frac{3(g_{\rm SM}-ge^{i\xi})}{v}\frac{HH}{2}t_L^\dagger t_R+{\rm h.c.},
\end{eqnarray}
contributes to the weak boson fusion process 
\begin{eqnarray}
    W^-(q,0)W^+(\bar{q},0)\to t\bar{t}H.
    \label{proc:WWttH}
\end{eqnarray}
The amplitudes in the SMEFT framework should hence be 
\begin{eqnarray}
    {\cal M}(W_L^-W_L^+\to t\bar{t}H)={\cal M}_{\rm complex Yukawa}+{\cal M}_{ttHH}.
\end{eqnarray}
With straightforward calculation, we obtain the analytical expression 
\begin{eqnarray}
{\cal M}_{ttHH}
=\frac{3}{v^2}
\left[
\mp 2p_t (g_{\rm SM}-g\cos\xi)
-i m_{tt} g\sin\xi
\right]
\frac{\hat{s}-2m_W^2}{\hat{s}-m_H^2},
\end{eqnarray}
when the $t$ and $\bar{t}$ have the same helicity $\pm\frac{1}{2}$ in the $t\bar{t}$ rest frame. The Goldstone boson equivalence theorem (GBET)~\cite{Cornwall:1974km,
Chanowitz:1985hj} tells that the amplitudes for the process~\eqref{proc:WWttH} should approach to those of the process 
\begin{eqnarray}
    \pi^-\pi^+\to t\bar{t}H.
\end{eqnarray}
From the Lagrangian~\eqref{eq:LSMEFT}, we can tell that the amplitudes are dominated by the contact $\pi^+\pi^-t\bar{t}H$ term, which gives  
\begin{eqnarray}
{\cal M}_{\pi\pi ttH}^{\pm\pm} =
\frac{1}{v^2}
\left[
\mp 2p_t(g_{\rm SM}-g\cos\xi)-im_{tt}g\sin\xi
\right].
\end{eqnarray}
The above two equations tell 
\begin{eqnarray}
    {\cal M}_{ttHH}\approx3{\cal M}_{\pi\pi ttH},
\end{eqnarray}
whereas the GBET tells, 
\begin{eqnarray}
    {\cal M}_{\rm complex Yukawa}+{\cal M}_{ttHH} \approx {\cal M}_{\pi\pi ttH},
\end{eqnarray}
at high energies.
Therefore it is straightforward to obtain the relation with high energy approximation.
\begin{eqnarray}
    {\cal M}_{\rm complex Yukawa}
    \approx-2{\cal M}_{\pi\pi ttH}.
\end{eqnarray}
The above relation for the scattering amplitudes ${\cal M} (W_L^-W_L^+\to ttH)$ explains the factor of 4 numerical difference of the total cross section in the complex Yuakwa model and the SMEFT model as shown in Figure~\ref{fig-1}(a) and~\ref{fig-1}(b).

\begin{figure}[t]
\centering
\subfigure[]{
\includegraphics[width=5cm,clip]{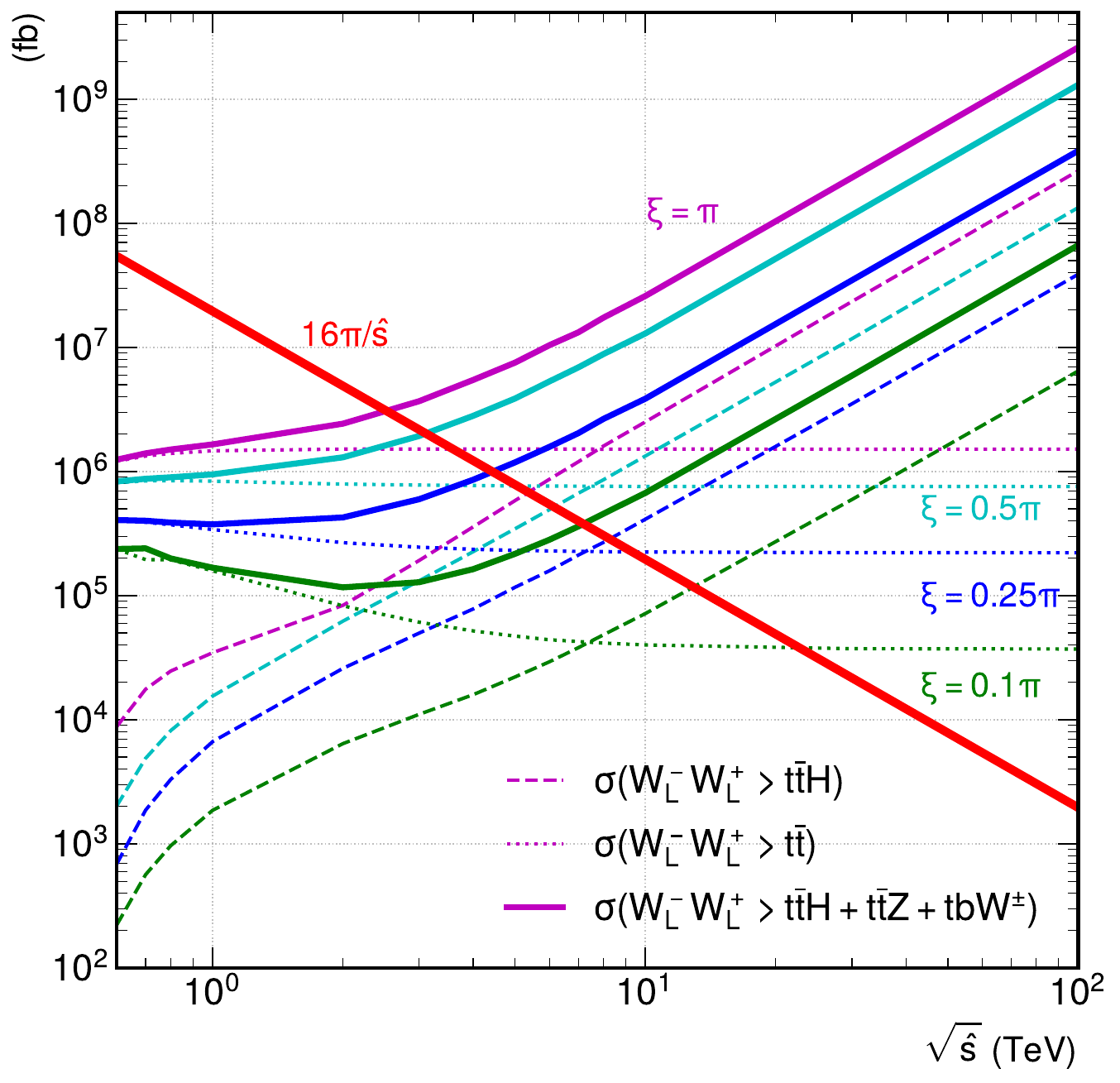}}
\subfigure[]{
\includegraphics[width=5cm,clip]{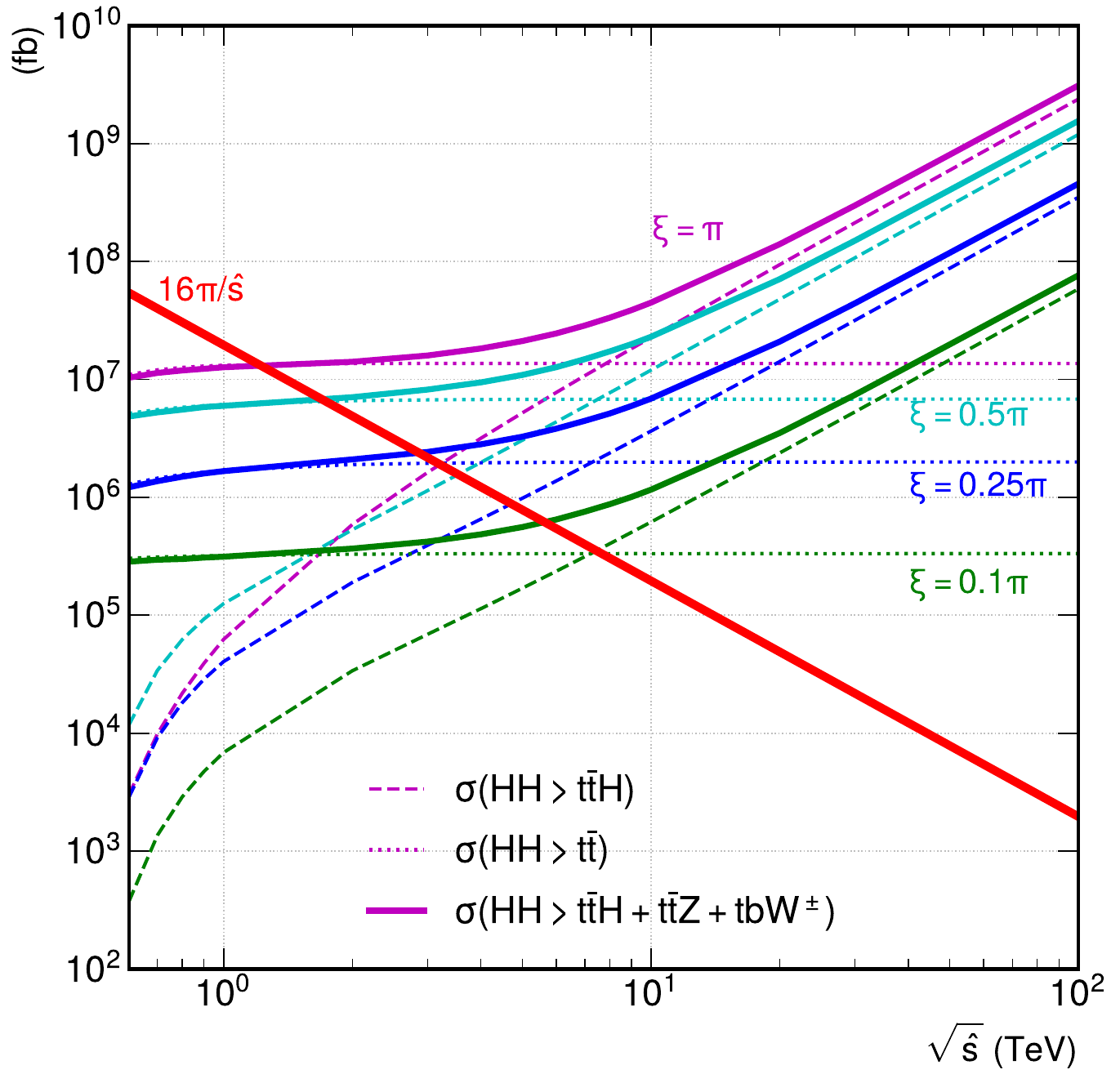}}
\caption{Unitarity bound from $W^-_LW^+_L$ and $HH$ channel summing over $2\to 2$ and $2\to 3$ processes.}
\label{fig-2}       
\end{figure}
The perturbative unitarity constraints of the SMEFT model can be obtained by the $J=0$ amplitudes of the weak boson fusion process. In the optical theorem, for the $W_L^-W_L^+$ process, the final state is summed over all $J=0$ final states,
\begin{eqnarray}
    2{\rm Im}\bra{i}T\ket{i}=\sum_f|\bra{f}T\ket{i}|^2,
\end{eqnarray}
with initial state $\ket{i}=\ket{W_L^-W_L^+(J=0)}$. The unitarity bound 
\begin{eqnarray}
    |{\rm Im}\bra{i}T\ket{i}|<|\bra{i}T\ket{i}|<16\pi
\end{eqnarray}
becomes
\begin{eqnarray}
    \sum_f\sigma_{\rm tot}(W_L^-W_L^+\to f;J=0)<\frac{16\pi}{\hat{s}}.
\end{eqnarray}

However, from the SMEFT Lagrangian, the large $ttHH$ and $ttHHH$ couplings make the $HH$ channel grows faster than the weak boson fusion channel. We show in Figure \ref{fig-2} for both cases. 
We note that the cases above the red line $\frac{16\pi}{\hat{s}}$ violating unitarity bound are non physical. The bounds are obtained from amplitudes of $\Lambda^{-2}$ order, since the $J=0$ amplitudes at high energies grow with energies by the dimension-5 and 6 vertices.

In summary, we investigate the strong energy dependence of the cross section for process $\mu^-\mu^+\to \nu_\mu\bar{\nu}_\mu t\bar{t}H$ in the complex top Yukawa model as well as the SMEFT framework. By comparing the two frameworks, we find that the highest-energy cross-section is reduced to one-quarter of the result predicted by the complex top Yukawa model, while maintaining the same energy dependence in the SMEFT with one dimension-6 operator\,\eqref{eq:LSMEFT}. This is shown explicitly by the dominant subamplitude for $W_L^-W_L^+\to t\bar{t}H$ by using the GBET.  The perturbative unitarity constraints are also obtained by summing up all $2\to 2$ and $2\to3 $ processes for $W_L^-W_L^+$ and $HH$ scattering ampiltudes.

\section*{Acknowledgement}
I extend my sincere thanks to the authors of the references\,\cite{Barger:2023wbg,Cassidy:2023lwd}, V. Barger, M.E. Cassidy, Z. Dong, K. Hagiwara, K. Kong, I.M.Lewis, and Y. Zhang for their valuable collaboration on the relevant study. The work was supported in part by JSPS KAKENHI Grant No.\,21H01077 and 23K03403.

%
 \bibliography{d6} 
%
%
%
%

\end{document}